\documentclass[preprint,12pt]{elsarticle}

\usepackage{amssymb}

\usepackage{color}
\usepackage{hyperref}
\usepackage{amsmath}

\biboptions{sort&compress}

\journal{Physica A}

\begin{document}

\begin{frontmatter}



\title{Collective behavior of coupled nonuniform stochastic oscillators}


\author[uefs,cbpf,ufpe]{Vladimir R. V. Assis}
\author[ufpe]{Mauro Copelli\corref{cor1}}
\cortext[cor1]{mcopelli@df.ufpe.br}
\address[uefs]{Departamento de F{\'\i}sica, Universidade Estadual de Feira de Santana, CEP 44031-460, Feira de Santana, Bahia, Brazil}
\address[cbpf]{Centro Brasileiro de Pesquisas F{\'\i}sicas, R. Dr. Xavier Sigaud, 150, 22290-180 Rio de Janeiro-RJ, Brazil}
\address[ufpe]{Departamento de F{\'\i}sica, Universidade Federal de Pernambuco, 50670-901 Recife, PE, Brazil}

\begin{abstract}
  Theoretical studies of synchronization are usually based on models
  of coupled phase oscillators which, when isolated, have constant
  angular frequency. Stochastic discrete versions of these uniform
  oscillators have also appeared in the literature, with equal
  transition rates among the states. Here we start from the model
  recently introduced by Wood {\it et al.\/} [Phys. Rev. Lett. {\bf
    96}, 145701 (2006)], which has a collectively synchronized phase,
  and parametrically modify the phase-coupled oscillators to render
  them (stochastically) nonuniform. We show that, depending on the
  nonuniformity parameter $0\leq \alpha \leq 1$, a mean field analysis
  predicts the occurrence of several phase transitions. In particular,
  the phase with collective oscillations is stable for the complete
  graph only for $\alpha \leq \alpha^\prime < 1$. At $\alpha=1$ the
  oscillators become excitable elements and the system has an
  absorbing state. In the excitable regime, no collective oscillations
  were found in the model.
\end{abstract}


\end{frontmatter}



\section{\label{intro}Introduction}

The last decades have witnessed the growth of a rich literature aimed
at developing theoretical tools for understanding the problem of
collective oscillatory behavior (often loosely termed
``synchronization'') emerging from the interaction of intrinsically
oscillating
units~\cite{Winfree67,Kuramoto84,Strogatz93,Strogatz00}. This
development has been prompted by the ubiquity of the phenomenon across
different knowledge areas, with abundant experimental
evidence~\cite{Huygens,StrogatzSync,Uhlhaas09}. Neuroscience is one
particular field where this problem reaches the frontiers of our
current theoretical understanding: neurons are highly nonlinear,
interact in large numbers and are often subject to
noise~\cite{Koch}. Under these conditions, it is not surprising that
many theoretical investigations on collective neuronal phenomena (of
which global oscillations is just a particular case) have been
restricted to numerical simulations~(recent examples include
Refs.~\cite{Ribeiro08a,Erichsen08,Agnes10}).

Analytical solutions of the synchronization problem in the presence of
noise have recently appeared, but at the cost of drastic
simplifications of each unit. For instance, each stochastic oscillator
in the model introduced by Wood {\it et al.\/}~\cite{Wood06a,Wood06b}
is described by three states connected by transition rates, amounting
to a discretization of a phase (already a simplification in
itself~\cite{Kuramoto84,Strogatz00}). The underlying assumption is
that the details in the description of each oscillator should become
increasingly irrelevant as the number $N$ of interacting units
diverges. This idea was beautifully illustrated in
Refs.~\cite{Wood06a,Wood06b}, which confirmed earlier predictions
(derived from a field-theory-based renormalization group
analysis~\cite{Risler04,Risler05}) that a phase transition to a
globally oscillating state belongs to the XY universality class.

Here we would like to apply this level of modeling to address a
different problem. We are concerned with global oscillations emerging
from interacting units which, when isolated, are {\em excitable\/},
i.e. not intrinsically oscillatory. In a detailed description (say, a
nonlinear ordinary differential equation), an excitable unit has a
stable fixed point (the quiescent state) from which it departs to a
long excursion in phase space if sufficiently perturbed. It then
undergoes a refractory period during which it is insensitive to
further perturbations, before returning to rest. A reduction of each
unit to a three-state system is straightforward in this
case~\cite{Lindner04}: employing the parlance of epidemiology, each
individual stays susceptible (S) unless it gets infected (I) from some
other previously infected individual, after which it eventually
becomes recovered (R) and finally becomes susceptible (S) again at
some rate. The prototypical experimental example of global
oscillations of excitable units comes precisely from epidemiology,
which shows periodic incidence of some diseases~\cite{AndersonMay}
(even though a person in isolation will not be periodically ill).

Models employing such cyclic three-state excitable units have usually
been termed SIRS models. These can be further subdivided into two
categories: those with a discrete-time description (cellular automata)
and those where time is continuous (frequently referred to as
``interacting particle systems''~\cite{Liggett1985}). In discrete-time
models, global oscillations have been
reported~\cite{Kuperman01,Gade2005,Kinouchi06a,Rozenblit11} and
understood analytically~\cite{Girvan02}. In continuous-time models,
the situation is different: in the standard SIRS model with diffusive
coupling~\cite{Joo04b}, global oscillations seem not to be stable,
even in the favorable condition of global or random
coupling~\cite{Joo04b,Rozhnova09,Rozhnova09b,Rozhnova09c}.

From a formal point of view, the cyclic structure of the excitable
units prevents the use of an equilibrium description. Moreover, the
system has a unique absorbing state (all units quiescent), which
provides an interesting connection with an enormous class of well
studied problems related to directed percolation~\cite{Marro99}. So,
within the context of continuous-time models, it is natural to ask:
are there markovian excitable systems with an absorbing state
(i.e. without external drive) which can sustain global oscillations?

We previously attempted to answer this question by modifying the
infection rate of the SIRS model, which was rendered exponential
(instead of linear, as in the standard case) in the density of
infected neighbors~\cite{Assis09}. The intention was to mimic the
exponential rates of Wood {\it et al.\/} which, in a system of
phase-coupled stochastic oscillators, did generate stable collective
oscillations~\cite{Wood06a,Wood06b}. In our nonlinearly pulse-coupled
system of excitable units, however, this modification actually failed
to generate oscillations, but rather turned the phase transition to an
active state discontinuous~\cite{Assis09}.

Here we make another attempt, now transforming the model by Wood {\it
  et al.\/} so that their oscillating units become increasingly
nonuniform.  Differently from our previous attempt~\cite{Assis09},
units here remain phase-coupled, even in the limit where the
nonuniformity transforms the oscillators into excitable units. Only in
this limit does the system have an absorbing state, and the question
is whether sustained macroscopic oscillations survive this microscopic
parametric transformation.

The paper is organized as follows. In section~\ref{model} we introduce
the model. Section~\ref{results} brings our main results, which are
further discussed with our concluding remarks in
section~\ref{conclusion}.

\section{\label{model}Model}

\begin{figure}[!hb]
  \begin{center}
    \includegraphics[width=.75\textwidth,angle=0]{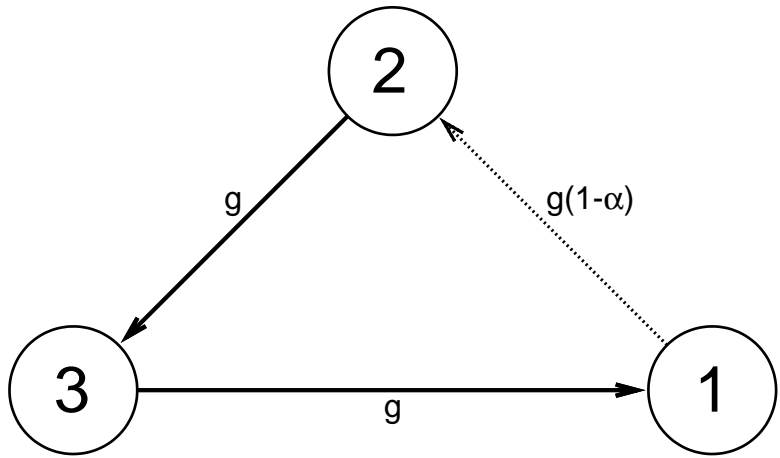}
    \caption{\label{fig:model}Graphic representation of the transition
      rates of the model for a single {\em isolated\/} unit.}
  \end{center}
\end{figure}

Let each unit $x$ ($x=1,\ldots,N$) be represented by a phase $\phi_x$,
which can be in one of three states: $\phi_x = (i_x-1)2\pi/3$, where
$i_x \in \{1,2,3\}$ (as illustrated in Fig.~\ref{fig:model}). When
isolated, the transition rates for each unit are
\begin{eqnarray}
\label{eq:rates.isolated.unit}
1 \longrightarrow 2 && \mbox{ at rate } g(1-\alpha), \nonumber \\
2 \longrightarrow 3 && \mbox{ at rate } g,  \\
3 \longrightarrow 1 && \mbox{ at rate } g \nonumber \; .
\end{eqnarray}
Parameter $\alpha$ controls the (average) nonuniformity of the
uncoupled oscillators. For $\alpha=0$ we recover the uniform
oscillators employed by Wood {\it et
  al.\/}~\cite{Wood06a,Wood06b}. For $0<\alpha<1$, however, each
oscillator tends to spend more time in state 1 than in the other
states, in what would be a stochastic version of a nonuniform
oscillation (such as, for example, that of an overdamped pendulum
under the effects of gravity and a constant applied
torque~\cite{Strogatz}). For $\alpha=1$, an {\em isolated\/} unit
stays in state 1 forever (though it will be able to leave this state
when coupled to other units, see below). The average angular frequency
of the uncoupled units is $\omega = 2 \pi g \left( 1 - \alpha \right)
/ \left( 3 - \alpha \right)$, which vanishes continually at
$\alpha=1$. This is qualitatively similar to what is observed in the
bifurcation scenario of type-I neuron
models~\cite{KochSegev}. Therefore, at $\alpha=1$ units become
excitable.

The coupling among units is essentially the same as the one used by
Wood~{\it et al.\/} in Refs.~\cite{Wood06a,Wood06b}, except that
parameter $\alpha$ controls nonuniformity in the transition from state
1 to state 2:
\begin{eqnarray}
  \label{eq:rates}
  1 \longrightarrow 2 & \mbox{ at rate } &
  g_{1,2}\left(\frac{N_1}{k},\frac{N_2}{k}\right) = g \left\{ e^{\left[a \left( N_2 - N_1 \right)\right]/k}
  - \alpha e^{\left(-a N_1\right)/k} \right\}, \nonumber \\
  2 \longrightarrow 3 & \mbox{ at rate } &
  g_{2,3}\left(\frac{N_2}{k},\frac{N_3}{k}\right) = g e^{\left[a \left( N_3 - N_2 \right)\right]/k}, \\
  3 \longrightarrow 1 & \mbox{ at rate } &
  g_{3,1}\left(\frac{N_3}{k},\frac{N_1}{k}\right) = g e^{\left[a \left( N_1 - N_3 \right)\right]/k}, \nonumber
\end{eqnarray}
where $g$ can be set to unity without loss of generality, $a$ is the
coupling parameter, $N_i$ is the number of neighbors in state $i$, and
$k$ is the total number of neighbors. Thus, for $\alpha=0$ the
original model by Wood {\it et al.\/} is recovered, whereas for
$\alpha=1$ the system has a single absorbing state (all units in
state~1).

\section{\label{results}Results}

\subsection{\label{subsec:MF}Mean field analysis}

\begin{figure}
\begin{center}
\includegraphics[width=0.7\textwidth,angle=0]{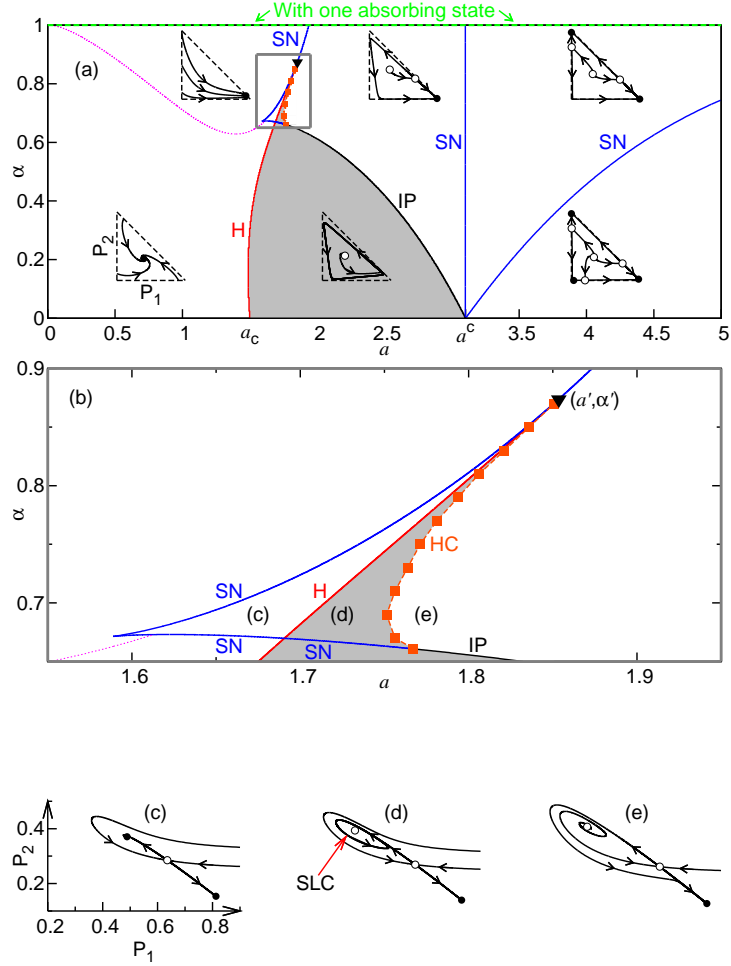}
\caption{\label{fig:phasediagram} (a) Phase diagram with phases
  characterized by representative phase portraits. The triangles mark
  the border of the physically acceptable region $0 \leqslant P_1
  \leqslant 1 $, $0 \leqslant P_2 \leqslant 1 $, and $0 \leqslant
  P_1+P_2 \leqslant 1$. Below (above) the dotted line [pink online]
  the stable fixed point is a spiral (node). The remaining lines
  represent bifurcations in the mean-field
  equations~\eqref{eq:MF.P1}-\eqref{eq:MF.P3}: Hopf (H) bifurcation
  (red online), saddle-node (SN) bifurcation (blue online) and
  infinite-period (IP) bifurcation (black). The squares indicate the
  homoclinic (HC) bifurcation line. The black triangle indicates the
  point~\mbox{$(a',\alpha')$} at which the HC, H and SN lines meet
  (see text for details). The dashed line (green online) on top of the
  panel~(at~\mbox{$\alpha=1$}) marks the parameter space region where
  there is a single absorbing state (but not necessarily a single
  attractor, see text for details).  (b)~Zoom of panel~(a) displaying
  details near the homoclinic line.  Panels~(c),~(d) and~(e) show the
  phase portraits for the parameters marked in the panel~(b), see also
  Fig.~\ref{P3F5}. SLC = stable limit cycle. There are stable limit
  cycles only in the grey regions.}
\end{center}
\end{figure}
Let~$g_{i,j}$ be the transition rate from state $i$ to state $j$,
given by~\eqref{eq:rates}, where~\mbox{$i \mbox{ and } j \in
  \{1,2,3\}$}. In a mean-field approximation, this leads to the
following equations:
\begin{eqnarray}
  \label{eq:MF.P1}
  \dot P_1 & = & g_{3,1}(P_3,P_1)P_3-g_{1,2}(P_1,P_2)P_1, \\
  \label{eq:MF.P2}
  \dot P_2 & = & g_{1,2}(P_1,P_2)P_1-g_{2,3}(P_2,P_3)P_2, \\
  \label{eq:MF.P3}
  \dot P_3 & = & g_{2,3}(P_2,P_3)P_2-g_{3,1}(P_3,P_1)P_3.
\end{eqnarray}
This also coincides with the equations for a complete graph in the
limit $N\to\infty$ with an inherent assumption that we can
replace~$N_i/N$ with~$P_i$.  Normalization~(\mbox{$P_3 = 1 - P_1 -
  P_2$}) reduces these equations to a two-dimensional flow in
the~$(P_1,P_2)$ plane.

We start by analyzing the phase diagram of the mean-field
equations~\eqref{eq:MF.P1}-\eqref{eq:MF.P3} in the phase
plane~\mbox{$(P_1,P_2)$}, which is restricted to the triangle \mbox{$0
  \leqslant P_1\leqslant 1$}, \mbox{$0 \leqslant P_2 \leqslant 1$},
and~\mbox{$0 \leqslant P_1+P_2 \leqslant 1$}. As shown in
Fig.~\ref{fig:phasediagram}, the case of uniform oscillators (where
the system has a perfect~$C_3$ symmetry) shows two bifurcations as one
increases $a$ on the line $\alpha=0$. First~\cite{Wood06a,Wood06b}, at
$a=a_c=1.5$, a supercritical Hopf bifurcation occurs above which the
symmetric fixed point $(P_1^*,P_2^*)=(1/3,1/3)$ loses stability and
stable collective oscillations emerge [a limit cycle in the phase
  plane $(P_1,P_2)$]. Increasing $a$ further, the frequency of these
global oscillations decreases continuously as the size of the limit
cycle increases~\cite{Wood07a,Wood07b}. For large values of $a$, the
shape of the limit cycle becomes less circular, approaching the
borders of the triangle, and global oscillations become highly
anharmonic, with a finite fraction of the oscillators collectively
spending a long time in each of the three states before ``jumping'' to
the next state at a much shorter time scale. At~$a=a^c\simeq 3.102$,
three saddle-node~(SN) bifurcations occur simultaneously at the limit
cycle, corresponding to an infinite-period transition in which~$C_3$
symmetry is broken~(since three stable attractors are
created)~\cite{Assis11a}. As noted previously~\cite{Wood07a},
this second transition is somewhat artificial from the perspective of
more realistic models, for which one expects (and observes)
oscillators to lock at a coupling-independent frequency. However, it
is very interesting from the perspective of nonequilibrium phase
transitions of interacting particle systems, since it provides a
spontaneous breaking of~$C_3$ symmetry in the absence of any absorbing
state~\cite{Assis11a} (as opposed to models with 3 absorbing
states, like rock-paper-scissors
games~\cite{Tainaka88,Tainaka89,Tainaka91,Tainaka94,Itoh94}).

For $\alpha > 0$, $C_3$ symmetry is no longer present. In fact, even
for arbitrarily small nonzero $\alpha$, the Hopf bifurcation is
followed (as $a$ is increased) by an infinite-period transition which
occurs due to a single SN bifurcation [as opposed to the triple SN for
  $\alpha=0$, see Fig.~\ref{fig:phasediagram}(a)]. Above this point,
collective oscillations disappear because units tend to condensate in
state 1, which is favored by the model nonuniformity [see
  Fig.~\ref{fig:model}]. Increasing $a$ further, another SN
bifurcation occurs [Fig.~\ref{fig:phasediagram}(a)], in which a value
of $P_2^*>1/3$ becomes a second stable fixed point of the model (i.e.,
state 2, which comes right after state 1, now can also attract the
oscillators). Finally, for even larger values of $a$, a third SN
bifurcation occurs and $P_3^*>1/3$ becomes a third attractor. With
three attractors, the phase space is similar to that observed for
$\alpha=0$.

Let us now address the stronger effects of the model
nonuniformity. Consider first the leftmost portion of
Fig.~\ref{fig:phasediagram}(a), i.e. for values of the coupling $a$
which are too small to yield sustained collective oscillations. In
this regime, the only attractor is a stable spiral, which for
$\alpha=0$ is symmetric: $(P_1^*,P_2^*)=(1/3,1/3)$. Increasing
$\alpha$, this spiral moves towards larger values of
$P_1^*$. Eventually, the imaginary part of its eigenvalues vanish and
the fixed point becomes a node [dotted line (pink online) in
Fig.~\ref{fig:phasediagram}a]. Only when $\alpha=1$ one reaches
$P_1^*=1$, which is then (and only then) an absorbing state
[Fig.~\ref{fig:phasediagram}a, upper dashed line (green online)].

Note that the infinite-period (IP) bifurcation line ends in the
beginning of the~SN and homoclinic bifurcation
lines~[Fig.~\ref{fig:phasediagram}(b)]. As we will see below, this
happens because there is an intrinsic interdependence among these
three bifurcations. An infinite-period bifurcation is just a~SN
bifurcation in which the saddle and the node are born at the limit
cycle, whereas in the homoclinic bifurcation the limit cycle collides
with a pre-existing saddle and becomes a homoclinic orbit. In both
cases the period diverges at the bifurcation, though with different
scaling regimes~\cite{Strogatz}.

Now we turn to the main question of this article, namely, what happens
if we start from a regime with collective oscillations~(say, with
\mbox{$a>a_c$},~\mbox{$\alpha\gtrsim 0$}) and increase the system
nonuniformity? We will see that many scenarios exist, but all of them
ultimately lead to the destruction of the limit cycle for
some~\mbox{$\alpha\leqslant\alpha'<1$}~[Fig.~\ref{fig:phasediagram}(b)].
We consider first the simplest cases. If, on one hand, we fix a value
of~$a$ very close to $a_c$, the size of limit cycle will continuously
decrease with increasing~$\alpha$ to vanish in a supercritical Hopf
bifurcation [H line (red online) in Figs.~\ref{fig:phasediagram}(a)
and~\ref{fig:phasediagram}(b)]. If the fixed value of~$a$ is
sufficiently close to~$a^c$, on the other hand, the limit cycle will
be destroyed by increasing~$\alpha$ via an infinite-period bifurcation
[IP line in Figs.~\ref{fig:phasediagram}(a)
and~\ref{fig:phasediagram}(b)].

There are intermediate values of~$a$ for which the way to the
annihilation of the limit cycle is more
tortuous~[Fig.~\ref{fig:phasediagram}(b)]. The key player to watch in
these scenarios is a point $(a^\prime,\alpha^\prime)$ (with
$a_c<a'<a^c$ and $0<\alpha'<1$) in parameter space where the Hopf
bifurcation line ends~[represented by a black triangle in the
  Figs.~\ref{fig:phasediagram}(a) and~\ref{fig:phasediagram}(b)]. For
any~$a\in(a_c,a')$, increasing~$\alpha$ will lead to extinction of the
limit cycle via a Hopf bifurcation~[Figs.~\ref{fig:phasediagram}(a)
  and~(b)]. For~\mbox{$a\lessapprox a'$}, the limit cycle is first
destroyed by an infinite-period bifurcation, then emerges again
through a homoclinic bifurcation, and finally disappears in a Hopf
bifurcation~[Fig.~\ref{fig:phasediagram}(b)]. There is also a small
reentrance in the homoclinic bifurcation
curve~[Fig.~\ref{fig:phasediagram}(b)], so that for any value of~$a$
in this short interval, the growth of~$\alpha$ leads to destruction
and recreation of limit cycle by two consecutive homoclinic
bifurcations.  Finally, for smaller values of~$a$, a~SN bifurcation
leads to the phase portrait of the Fig.~\ref{fig:phasediagram}(d)
before the limit cycle is annihilated by a Hopf
bifurcation~[Fig.~\ref{fig:phasediagram}(b)].

If we set~$\alpha > \alpha^\prime$ and $a<a^\prime$, we will have only
one stable node with~$P_1^*\approx 1$ (remember that for $\alpha=1$ we
have~$P_1^*=1$, which corresponds to the absorbing state). Increasing
$a$, there will be a first SN bifurcation at which a saddle is born
with an {\em unstable\/} node (which quickly becomes a
spiral)~[Fig.~\ref{fig:phasediagram}(a)]. Further increases in $a$
will give rise to two additional SN bifurcations, as described
above. After the third SN
bifurcation, the system will have three attractors with a phase
portrait of the same type as that of the
figure~\ref{P3F4}(b). Importantly, besides the three saddle-node
bifurcations mentioned, no other bifurcation occurs
for~$\alpha>\alpha'$. In particular, we found no limit cycles for
$\alpha^\prime < \alpha < 1$, so that the nonuniformity alone can
destroy collective oscillations, even in the absence of an absorbing
state. 

\subsection{Complete graph simulations}

\begin{figure}[!ht]%
  \begin{center}%
    \includegraphics[width=1.\columnwidth,angle=0]{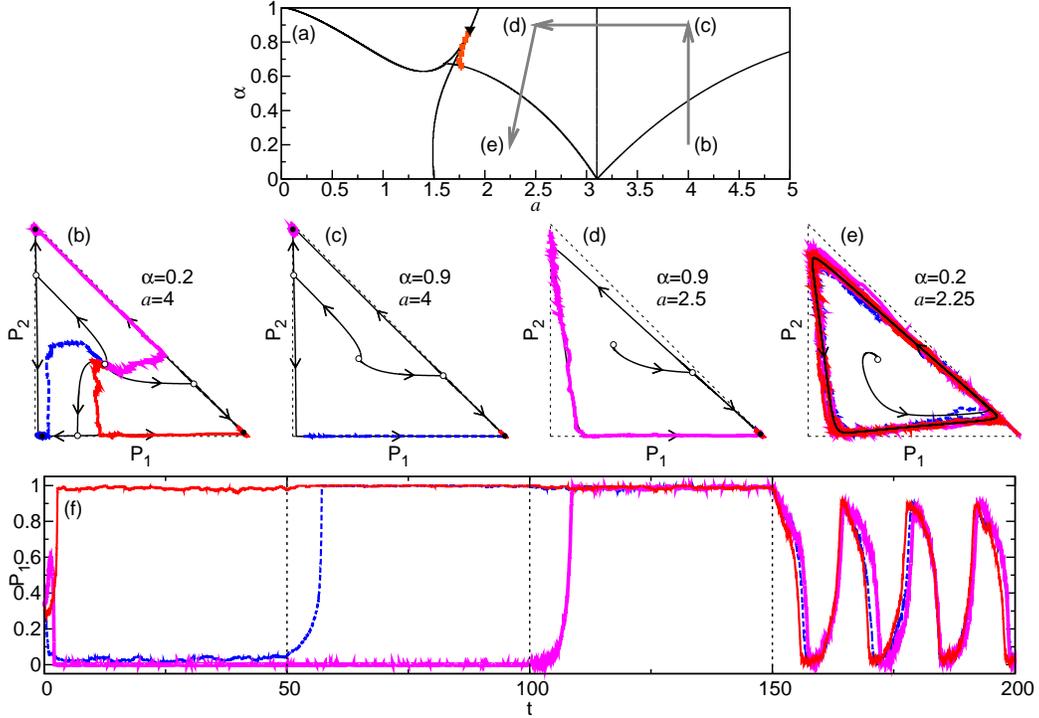}%
    \caption{\label{P3F4} Phase portraits and simulation results for a
      complete graph with $N=500$. Every 50 time units, the parameters
      of the model are changed so as to eliminate a stable fixed point
      of the model. (a)~Phase diagram with labels to indicate the
      parameter-space places of the phase portraits (b)-(e) and arrows
      to show the order of the parameter-value jumps. (b)~Phase
      portrait for $\alpha=0.2$ and $a=4$. The three samples of the
      simulations begin at the unstable fixed point close to triangle
      center, then each one converges to a different attractor owing
      to fluctuations.  (c)~Phase portrait for $\alpha=0.9$ and
      $a=4$. During the simulations, we change the parameter $\alpha$
      to the value of this phase portrait.  Thus, one of the
      attractors disappears, so the sample which was around one
      attractor converges to the stable fixed point of the attraction
      basin where this sample is now. Obviously, the other two samples
      continue around their attractor. (d)~Phase portrait for
      $\alpha=0.9$ and $a=2.5$. Similarly to panel (c), another
      attractor disappears. Thus, the sample which was isolated
      converges to the remaining stable fixed point. (e)~Phase
      portrait for $\alpha=0.2$ and $a=2.25$. Finally, the remaining
      stable fixed point disappears, so the three samples oscillate
      almost together (except for fluctuations). Panel (f) shows time
      series for simulated trajectories shown in panels (b), (c), (d),
      and (e), in chronological order. The false impression that the
      inequalities $0\leq P_j \leq 1$ are sometimes violated is
      caused by the thickness of the lines.}
  \end{center}%
\end{figure}%
As mentioned in subsection~\ref{subsec:MF}, the mean-field
calculations are exact for the complete graph in the thermodynamic
limit. For completeness, we now study the effects of finite-size
fluctuations. While we do not expect stable collective oscillations to
appear for $\alpha\geqslant\alpha^\prime$, stochastic oscillations
have been predicted to appear around (mean-field) stable
spirals~\cite{Risau-Gusman07} in finite populations. Moreover,
fluctuations become particularly prominent in systems with
multistability, ultimately determining the attractor to which the
system will converge.

\begin{figure}[!hb]%
  \begin{center}%
    \includegraphics[width=0.7\columnwidth,angle=0]{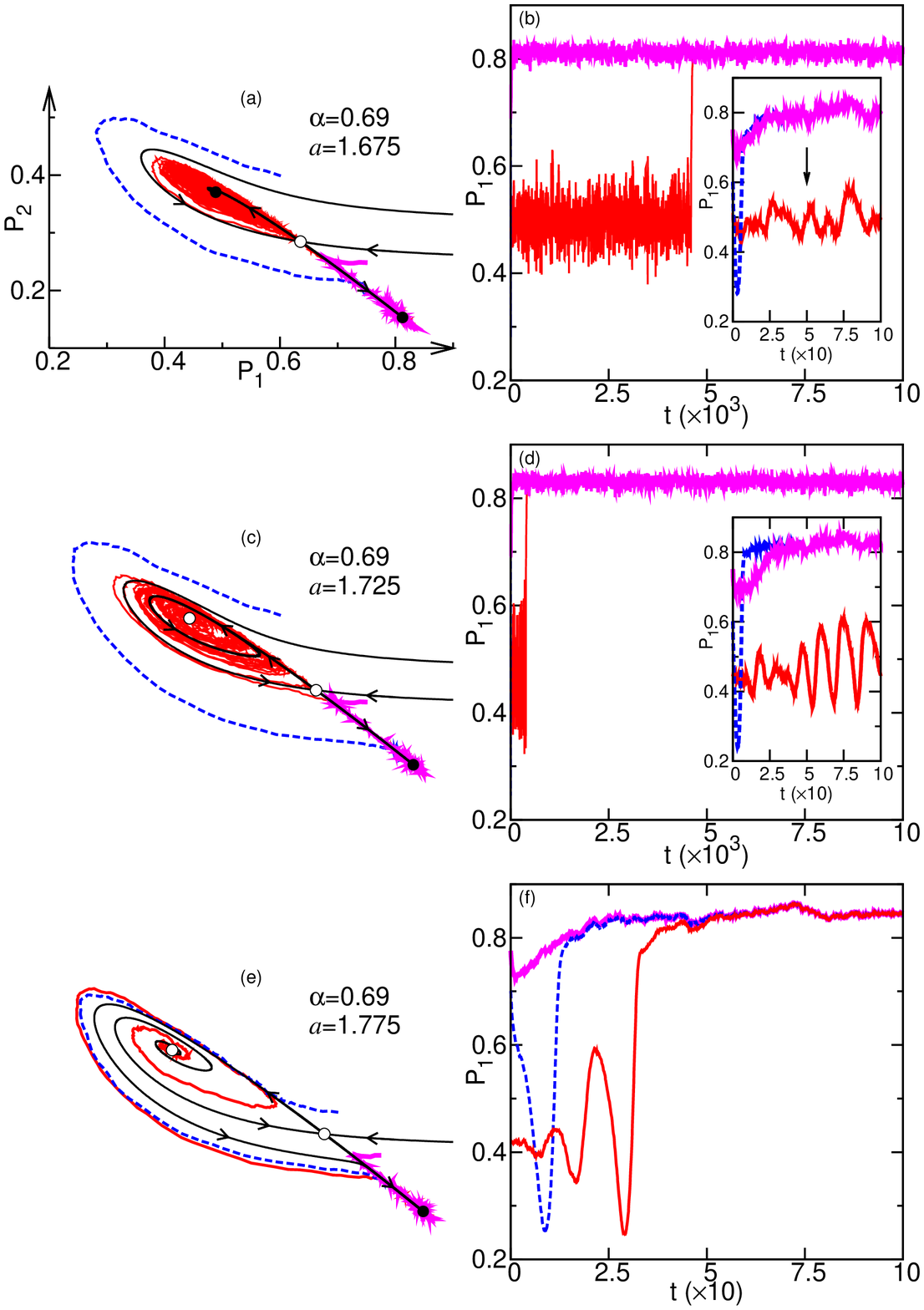}%
    \caption{\label{P3F5} Phase portraits and simulation results for a
      complete graph with $N=20000$, see
      Fig. \ref{fig:phasediagram}. (a) Phase portrait for
      $\alpha=0.69$ and $a=1.675$. There are two attractors: a node
      and a spiral (they are separated by a saddle). (c) Phase
      portrait for $\alpha=0.69$ and $a=1.725$. The spiral becomes
      unstable via a Hopf bifurcation and is surrounded by a stable
      limit cycle. (e) Phase portrait for $\alpha=0.69$ and
      $a=1.775$. After a homoclinic bifurcation, the limit cycle
      disappears.  Panels (b), (d), and (f) show time series for
      simulated trajectories, respectively, shown in panels (a), (c),
      and (e), with examples of collective excitability (dashed, blue
      online) and stochastic oscillations [arrow in the inset of
      (b)].}%
  \end{center}%
\end{figure}%

First, let us deal with the effects of finite-size fluctuations in a
scenario with multistability by starting the system near an unstable
stationary solution at the intersection of the basins of attraction of
the three stable solutions~[Fig.~\ref{P3F4}(b)]. At point (b) in
Fig.~\ref{P3F4}(a), tristability is revealed by the different fates of
the system for three independent realizations of the same macroscopic
initial condition.

Next, after each sample was already in its respective steady state, we
have increased~$\alpha$ discontinuously [(b)$\rightarrow$(c) in
Fig.~\ref{P3F4}(a)] without interruption in the simulations in order
to extinguish one of the attractors. Therefore, the sample which was
around the newly extinct attractor converges to the stable solution of
the new larger basin of attraction [while the other two samples are
only slightly disturbed, see~Fig.~\ref{P3F4}(c)].

Then, similarly to the previous case, we decrease~$a$ discontinuously
[(c)$\rightarrow$(d) in Fig.~\ref{P3F4}(a)] in order to suppress the
attractor of the isolated sample, which is forced to converge to the
remaining stable fixed point. At this stage, all three samples have
the same behavior, apart from fluctuations~[Fig.~\ref{P3F4}(d)].

Finally, we reduce~$a$ and~$\alpha$ discontinuously and simultaneously
[(d)$\rightarrow$(e) in Fig.~\ref{P3F4}(a)] in order to eliminate the
last stable fixed point. The three samples thereafter converge (at
about the same time) to a stable limit cycle in which they oscillate
almost together, apart from fluctuations~[Fig.~\ref{P3F4}(e)
and~(f)]. Note that the process illustrated in Figure~\ref{P3F4} does
not close a cycle and is irreversible: no matter whether we change the
control parameters to their initial values [i.e. (e)$\rightarrow$(b)
in Fig.~\ref{P3F4}(a)], or we apply the reverse changes
[(e)$\rightarrow$(d)$\rightarrow$(c)$\rightarrow$(b)], it is extremely
unlikely that the system will return to its initial steady state
[Figure~\ref{P3F4}(b)].

Now consider the phase diagram of Figure~\ref{fig:phasediagram}(b). In
the far left (lower values of~$a$) there is only one stable
node. Increasing~$a$, a SN bifurcation occurs in which a newly arisen
node turns into a spiral after an extremely small increase
in~$a$. Thus, we get the phase portrait of the
Figure~\ref{fig:phasediagram}(c), whose simulation results for a
complete graph appear in Figures~\ref{P3F5}(a) and~(b). Although the
mean-field calculations predict that the spiral in
Figures~\ref{fig:phasediagram}(c) and~\ref{P3F5}(a) is stable, even
for large complete graphs ($N=20000$ in Figure~\ref{P3F5}) finite-size
fluctuations eventually lead to system to the other (fixed point)
attractor. However, before that happens, these fluctuations lead to
stochastic oscillations around the spiral~\cite{Risau-Gusman07}~[see
arrow in the inset of Figure~\ref{P3F5}(b)].

Increasing~$a$ further, the spiral loses stability in a Hopf
bifurcation, giving rise to a stable limit cycle~(SLC). We arrive then
at the phase portrait shown in Figure~\ref{fig:phasediagram}(d),
with simulation results for a complete graph shown in
Figures~\ref{P3F5}(c) and~(d). Again, while the mean-field
calculations predict that the limit cycle in
Figures~\ref{fig:phasediagram}(d) and~\ref{P3F5}(c) is \mbox{\em
  stable\/}, finite-size fluctuations lead the system to the other
attractor in a considerably shorter time than in the previous
case~[Figs.~\ref{fig:phasediagram}(c),~\ref{P3F5}(a) and~(b)].

Soon after the Hopf bifurcation, increasing~$a$ leads to an increase
in the size of the~SLC, until it is finally destroyed by a homoclinic
bifurcation~[Fig.~\ref{fig:phasediagram}(b)]. Thus, we reach the phase
portrait of the Figure~\ref{fig:phasediagram}(e), with the results of
simulations of a complete graph shown in Figures~\ref{P3F5}(e)
and~(f). Now there is only one attractor in the system. As expected,
if we start the system in the unstable spiral, there will be
oscillations of increasing amplitude until the system reaches the
stable steady state.

Note that the presence of the saddle in
Figs.~\ref{fig:phasediagram}(c)-(e) sets the conditions for {\em
  collective excitability\/}~\cite{Assis09}. If the system is
initially in the stable node, perturbations that throw the system to a
different point still within its basin of attraction can lead to
qualitatively different relaxation processes. Small perturbations
decay monotonically, whereas large perturbations will take the system
further away from the fixed point before returning to it (because the
system is required to go around the stable manifold of the
saddle). This can be clearly seen in the simulations (compare the
dashed and thick lines --- blue and pink online, respectively --- in
Figure~\ref{P3F5}).

\section{\label{conclusion}Concluding remarks}

We have proposed a variant of the model by Wood {\it et
  al.\/}~\cite{Wood06a,Wood06b} in which stochastic oscillators can
become increasingly nonuniform as parameter $\alpha$ goes from 0 to
1. We have obtained the phase diagram of the mean-field version of the
model in the $(a,\alpha)$ parameter plane. In addition to the
previously reported phase transitions for
$\alpha=0$~\cite{Wood06a,Wood06b,Assis11a}, we have obtained for
$\alpha\neq 0$ several bifurcations in the mean-field equations of the
model, including saddle-node, infinite period, Hopf and
homoclinic. Collective excitability~\cite{Assis09} has been shown to
occur in some parameter regions, as confirmed by simulations of
complete graphs. Simulations have also confirmed the overall
predictions of the mean-field analysis, although the stability of some
stable limit cycles and fixed points has failed to resist the effects
of finite-size fluctuations.

In the regime in which the units are excitable elements, we did not
find stable global oscillations for the particular choice of a
nonlinear coupling defined by Eqs.~\ref{eq:rates}, even in the
complete graph. This topology is the one in which a synchronized
solution would be most favorable, as shown in a number of previous
reports~\cite{Kuperman01,Wood06a,Wood06b,Risau-Gusman07,Wood07a,Wood07b},
which retrospectively justifies why we have not attempted to run
simulations of the model in a hypercubic (or even small-world)
topology. We have shown that, in the model of Wood~{\it et
  al.\/}~\cite{Wood06a,Wood06b}, nonuniformity hinders the
synchronization among the oscillators, while large enough
nonuniformity and, consequently, the transformation of oscillatory
units into excitable elements prevents the emergence of a synchronized
stable state. It remains to be investigated whether this can be
achieved with a different type of coupling.

Our results nonetheless raise some interesting questions which deserve
further investigation. For any nonuniformity ($\alpha > 0$) the
transition to a synchronized state in the complete graph occurs in the
absence of $C_3$ symmetry. Does the transition still occur in
hypercubic lattices under these conditions~\cite{Wood06a}? If so, do
the critical exponents depend on $\alpha$? Investigations in these
directions would certainly be welcome.

\section{Acknowledgments}
VRVA and MC acknowledge financial support from CNPq, FACEPE, CAPES,
FAPERJ and special programs PRONEX, PRONEM and INCEMAQ.

\bibliography{copelli-local}





\bibliographystyle{model1-num-names}







\end{document}